\documentclass[a4paper]{article}
\makeatletter
\def\@seccntformat#1{\@ifundefined{#1@cntformat}%
   {\csname the#1\endcsname\quad}  
   {\csname #1@cntformat\endcsname}
}
\let\oldappendix\appendix 
\renewcommand\appendix{%
    \oldappendix
    \newcommand{\section@cntformat}{\appendixname~\thesection\quad}
}
\makeatother

\usepackage{graphicx}

\usepackage{cite}
\usepackage[normalem]{ulem}

\usepackage[usenames,dvipsnames]{xcolor} 

\usepackage{amssymb,amsfonts,amsmath}
\usepackage[top=25truemm,bottom=25truemm,left=20truemm,right=20truemm]{geometry}
\usepackage{bm}
\usepackage{cases}
\usepackage[mathlines]{lineno}

\usepackage{setspace}

\usepackage{url}
\usepackage{comment}

\title{How mutation alters the evolutionary dynamics of cooperation on networks}
\bigskip
\author{Genki Ichinose${}^{1*}$, Yoshiki Satotani${}^{2}$, Hiroki Sayama${}^{3}$
\ \\
\ \\
${}^{1}$
Department of Mathematical and Systems Engineering, Shizuoka University, \\3-5- 1 Johoku, Naka-ku, Hamamatsu, 432-8561, Japan\\
${}^{2}$
Department of Information Technology, Okayama University, Okayama 700-8530 Japan\\
${}^{3}$
Center for Collective Dynamics of Complex Systems\\
Department of Systems Science and Industrial Engineering\\ Binghamton University, State University of New York, Binghamton, NY 13902-6000\\
$^*$ Corresponding author (ichinose.genki@shizuoka.ac.jp)}

\begin{document}

\maketitle

\section*{Abstract}
Cooperation is ubiquitous at every level of living organisms.
It is known that spatial (network) structure is a viable mechanism for cooperation to evolve.
A recently proposed numerical metric, Average Gradient of Selection (AGoS), a useful tool for interpreting and visualizing evolutionary dynamics on networks, allows simulation results to be visualized on a one-dimensional phase space.
However, stochastic mutation of strategies was not considered in the analysis of AGoS.
Here we extend AGoS so that it can analyze the evolution of cooperation where mutation may alter strategies of individuals on networks.
We show that our extended AGoS correctly visualizes the final states of cooperation with mutation in the individual-based simulations.
Our analyses revealed that mutation always has a negative effect on the evolution of cooperation regardless of the payoff functions, fraction of cooperators, and network structures.
Moreover, we found that scale-free networks are the most vulnerable to mutation and thus the dynamics of cooperation are altered from bistability to coexistence on those networks, undergoing an imperfect pitchfork bifurcation.

\section*{Keywords}
Evolution of cooperation; Prisoner's dilemma game; Network structure; Average gradient of selection; Equilibrium point; Bifurcation

\section{Introduction\label{sec:introduction}}
Cooperation is ubiquitous at every level of living organisms and has played an important role in major evolutionary transitions \cite{West18082015, Michod15052007, Michod13062006}.
In principle, cooperators benefit others by incurring some costs to themselves, while defectors do not pay any costs.
Therefore, cooperation cannot be an evolutionarily stable strategy for a noniterative game in a well-mixed population \cite{nowak06evolutionaryDynamicsBOOK, MaynardSmith1973, TaylorJonker1978, MaynardSmithBOOK1982, HofbauerBOOK1998}.

In such a situation, spatial (network) structure is a viable mechanism for cooperation to evolve \cite{NowakMay1992} (Refer to the various models for cooperation on networks in the comprehensive reviews \cite{PercSzolnoki_Biosystems2010, Wang_etal_EJB2015, Perc_etal_PhysRep2017}).
In particular, cooperation is significantly enhanced on scale-free networks with accumulated payoffs \cite{SantosPacheco2005}. In this case, cooperative strategies can spread into the whole population because cooperative hubs surrounded by other cooperators obtain much higher payoffs compared to other nodes \cite{SantosPacheco2005}.
Due to the development of evolutionary graph theory \cite{Lieberman_etal2005}, mathematical understanding of the evolution of cooperation on networks has been greatly advanced \cite{Ohtsuki_etal2006, OhtsukiNowak2006, Taylor_etal2007, AllenNowak2014, Debarre_etal2014, Maciejewski_etal2014, Tarnita_etal2014, Allen_etal2017}. As a pioneering solid mathematical work, Ohtsuki et al. calculated the fixation probability of cooperators on various networks and revealed that the relationship between the benefit-cost ratio and the average degree simply determines whether cooperation can evolve on those networks \cite{Ohtsuki_etal2006}.
Although mathematical frameworks are a powerful tool to capture the whole structure of evolutionary games, some assumptions such as weak selection \cite{Ohtsuki_etal2006, OhtsukiNowak2006, Taylor_etal2007, AllenNowak2014, Debarre_etal2014, Maciejewski_etal2014, Tarnita_etal2014, Allen_etal2017} and homogeneous graphs (all nodes have the same degree) \cite{OhtsukiNowak2006, Taylor_etal2007, AllenNowak2014, Debarre_etal2014} are often associated with them.
To tackle this problem, numerical methods have some advantages because they allow complex situations such as non-weak selection and heterogeneous graphs to be calculated  once those parameters are given.
Pinheiro et al.~\cite{PinheiroPachecoSantos2012} proposed a notable numerical metric, called Average Gradient of Selection (AGoS).
AGoS can interpret and visualize the dynamics of simulation results in a one-dimensional phase space by utilizing a numerical mean-field method.
It can analyze the dynamics of the evolution of cooperation in structured populations (including heterogeneous graphs) even when non-weak selection pressure is introduced \cite{PinheiroSantosPacheco2012}, when nonlinear imitation probability is used \cite{Dai_etal2013}, and also when structures and states of networks change over time (adaptive social networks) \cite{PinheiroSantosPacheco2016}.

In these earlier studies of the AGoS, however, stochastic mutation of strategies was not considered.
It is important to incorporate such mutation because they frequently occur in real societies and also because results obtained with stochastic fluctuations of strategies would provide more robust observations and conclusions.
Prominent mathematical results of mutation effects on networked cooperation have been found \cite{Allen_etal2012, Allen_etal2017, AllenNowak2014, Tarnita_etal2014, Debarre_etal2017}. Especially, Allen et al. \cite{Allen_etal2012} and D\'ebarre \cite{Debarre_etal2017} deeply analyzed the effects of mutation on networked cooperation and found that mutation has a negative effect for cooperation by diluting the clustering of cooperators. In their models, homogeneous graphs and weak selection are assumed. Allen et al. \cite{Allen_etal2017} recently obtained elegant mathematical results of this problem under weak selection on any population structure including heterogeneous graphs.
Here we analyze the evolution of cooperation using AGoS where mutation may alter strategies of individuals in networks.
Compared to these previous AGoS and mathematical results, our contribution of this paper to the literature  is to use AGoS for the evolution of cooperation on both homogeneous and heterogeneous graphs with non-weak selection.
We also consider the Fermi update rule for the strategy updating, which is usually a different scheme than is used in mathematical studies.
On the other hand, statistical physics has also played an important role in the literature of cooperation on networks \cite{Perc_etal_PhysRep2017}. If the network reciprocity and some other mechanisms are combined, cooperation is further promoted, which is known as coevolutionary rules \cite{PercSzolnoki_Biosystems2010}. Examples of the coevolutionary rules are found in Refs.~\cite{SzolnokiPerc_NJP2008, PercSzolnoki_PRE2008, Wang_etal_PONE2012, Wang_etal_SciRep2012, Wang_etal_PRE2012, WangPerc_PRE2010, PercWang_PONE2010, Wang_etal_PRE2012b, Wang_etal_JTB2011, SzolnokiPerc_NJP2018, Battiston_etal_PRE2017, Aleta_etal_PRE2016, SzolnokiPerc_EPL2009, Perc_NJP2006, SzolnokiPerc_NJP2009}. Our model does not fall into the framework of the coevolutionary rules because there is no additional mechanism other than the network reciprocity.
However, our model is in line with the statistical physics of cooperation. Based on this framework, this paper shows that scale-free networks are quite vulnerable to mutation at a macroscopic level.

\section{Models}
We developed an agent-based model for the analysis of AGoS.
Individuals are placed on the nodes in a network and they interact only with their neighbors.
The networks used in this paper are described in the next subsection.
Each individual can take one of two strategies: Cooperation ($C$) or Defection ($D$).

Once the composition of individuals in a network is given, we can calculate the probability of increasing or decreasing the number of cooperators by one, called Gradient of Selection (GoS) at time $t$ \cite{PinheiroPachecoSantos2012}.
Simultaneously, we can also update the strategies of individuals based on the framework of evolutionary games at time $t$.
Immediately after the strategy updating, mutation (flipping strategy from $C$ to $D$ or vice versa) occurs with probability $m$.
Therefore, in evolutionary simulations, the calculations of GoS and the strategy updating with mutation take place alternately in one time step and these processes are repeated.
The calculation of GoS and the strategy updating with mutation are repeated for $\Lambda$ time steps.
Simulations are repeated for $\Omega$ times in total.

\subsection*{Network structure}
Pinheiro et al.~\cite{PinheiroPachecoSantos2012} revealed that, by using AGoS, cooperation was sustained at a network level and network structures led to the different evolutionary results of cooperation.
Following the existing work \cite{PinheiroPachecoSantos2012, PinheiroSantosPacheco2012}, we focus on two classes of network structures: homogeneous and heterogeneous.
We use homogeneous in the sense that every individual has the same degree. In the case of heterogeneous, individuals can have different degrees.

For homogeneous networks, we use two types: homogeneous random networks (HR) \cite{SantosRodriguesPacheco2005} and square lattice (Lattice). HR is created by randomizing links from homogeneous regular ring networks.
For heterogeneous networks, we use Barab\'asi-Albert scalfe-free networks (BA) \cite{BarabasiAlbert1999}. In BA, a small number of nodes called hubs connect with a substantial number of links while most other nodes connect with a few nodes.

In the initial setting, the number of cooperators in a population, $j \in [0, N]$, is given to each simulation. The locations of $j$ cooperators and $N-j$ defectors in a network are random. 

\subsection*{AGoS calculation}
Once each simulation starts, GoS of the given population at time $t$ is calculated.
If there was no population (network) structure, the dynamics of cooperation could be calculated analytically using the replicator equation \cite{HofbauerBOOK1998, nowak06evolutionaryDynamicsBOOK}.
However, in structured populations, the calculation becomes difficult due to the complex topologies of networks although many mathematical results with some assumptions have been obtained \cite{Ohtsuki_etal2006, OhtsukiNowak2006, Taylor_etal2007, AllenNowak2014, Debarre_etal2014, Maciejewski_etal2014, Tarnita_etal2014, Allen_etal2017}.

The GoS gives the numerical characterization of the evolution of cooperation even in such a situation.
Let $\phi_i (t)$ be the probability that $i$'s strategy changes to the other different strategy (from $C$ to $D$ or from $D$ to $C$) and $n_i$ is a set of $i$'s neighbors that have the different strategy from $i$'s.
Then, $\phi_i (t)$ can be the product of two terms: the probability of selecting a neighbor with the different strategy, $\frac{|n_i|}{k_i}$, and the average probability that $i$ imitates the different strategy, $\frac{1}{|n_i|} \sum_{l\in n_i} p(i, l, t)$, where $k_i$ is the number of $i$'s neighbors and $p(i, l, t)$ is the probability that individual $i$ imitates $l$'s strategy at time $t$, which is defined in the next subsection.
Thus, $\phi_i (t)$ can be written as follows:

\begin{equation}
\phi_i (t)=\frac{|n_i|}{k_i}\frac{1}{|n_i|}\sum_{l\in n_i} p (i, l, t)=\frac{1}{k_i}\sum_{l\in n_i} p (i, l, t)
\label{phi}
\end{equation}

From the definition, for a given simulation $r$, the probability to increase the number of $C$ at $j$ (the number of $C$s in the network) is $\phi_{r}^{+} (j,t)=\frac{1}{N}\sum_{i \in s_D}\phi_i (t)$ while the probability for the decrease of the number of $C$s is $\phi_{r}^{-}(j, t) =\frac{1}{N}\sum_{i \in s_C} \phi_i (t)$, where $N$ is the population size, $s_D$ is a set of defectors ($|s_D|=N-j$), and $s_C$ is a set of cooperators ($|s_C|=j$).
Then, GoS at time $t$ in a given simulation $r$ is defined as the difference between them, as

\begin{equation}
G_r (j, t)=\phi_{r}^{+}(j, t) - \phi_{r}^{-}(j, t).
\label{gos}
\end{equation}

Finally, we obtain the Average Gradient of Selection (AGoS) \cite{PinheiroPachecoSantos2012}, which averages the GoS by the number of time steps $\Lambda$ and the number of simulations $\Omega$, as

\begin{equation}
G^{A} (j)=\frac{1}{\Lambda \Omega}\sum_{r=1}^{\Omega}\sum_{t=1}^{\Lambda} G_r (j,t).
\label{agos}
\end{equation}

The above derivation is the original AGoS without mutation \cite{PinheiroPachecoSantos2012}.
Here we need to consider the effect of mutation on cooperation.
To incorporate mutation, Eq.~\eqref{phi} needs to be revised so that the changed strategy by selection remains at the same state with probability $1-m$ while the unchanged strategy by selection changes to the opposite strategy with mutation probability $m$.
Thus, the probability that $i$'s strategy changes to the other different strategy is altered to

\begin{equation}
\phi_i (t)=(1-m)\frac{1}{k_i}\sum_{l\in n_i} p (i, l, t)+m \left[1- \frac{1}{k_i}\sum_{l\in n_i} p (i, l, t)\right].
\label{phi_new}
\end{equation}

Note that Eq.~\eqref{phi_new} returns to Eq.~\eqref{phi} when $m=0$ (without mutation).
Instead of Eq.~\eqref{phi}, we use Eq.~\eqref{phi_new} for calculation of GoS (Eq.~\eqref{gos}) and AGoS (Eq.~\eqref{agos}).
We use these extended versions for our analyses.

\subsection*{Strategy updating with mutation}
To check the accuracy of the AGoS, we simultaneously conducted individual-based simulations.
After the GoS calculation, the strategy updating takes place as follows.
In each time step, we randomly choose one individual $i$ from the population.
The individual plays the Prisoner's Dilemma (PD) game with its $k_i$ neighbors and obtains payoffs resulting from the games.
In each game, both individuals obtain payoff $R$ for mutual cooperation or $P$ for mutual defection.
If one selects cooperation while the other defects, the former obtains the sucker's payoff $S$ while the latter obtains the highest payoff $T$, the temptation to defect.
The relationship of the four payoffs is usually $T > R > P > S$ in PD games.
AGoS is used for the other types of collective games including Stag Hunt \cite{Pacheco_etal2009} and Snowdrift \cite{Santos_etal2012} games.

Following the parameter settings used in the model by \cite{PinheiroPachecoSantos2012, PinheiroSantosPacheco2012}, we used $P = 0, R = 1$, and $S = 1-T$, while $T > 1$.
All the neighbors of $i$ also play the PD game with their neighbors and obtains the payoffs.
Let $\pi_i$ and $\pi_l$ be the payoffs of individual $i$ and $l$ (a randomly selected neighbor of $i$), respectively.
Here we consider the two types of payoff functions: accumulated payoff and average payoff, because results may be quite different between those two conditions. In the accumulated payoff, cooperation is greatly promoted due to the hub effect \cite{SantosPacheco2005}.
In the average payoff \cite{SantosPacheco2006, Szolnoki_etal2008, Tomassini_etal2007, Zhi-Xi_etal2007, IchinoseSayama2017} (or in the case that having links takes some costs \cite{Masuda2007}), however, cooperation is inhibited because the hub effect is canceled.
In the former case, the payoffs obtained from each game are accumulated over all the neighbors.
In contrast, in the latter case, the accumulated payoffs for $i$ at the end of the games are divided by degree $k_i$. 
In either case, the payoff is reset to zero at every time step.

Based on the framework of evolutionary games, higher fitness will be imitated more often.
In order to realize this, we use the pairwise comparison rule \cite{TraulsenNowakPacheco2006, TraulsenPachecoNowak2007, SzaboToke1998}.
As mentioned above, first, a target node $i$ (learner) is randomly selected from the population. Then, one neighbor $l$ (teacher) of the target node is randomly selected. Finally, individual $i$ imitates $l$'s strategy with the probability 

\begin{equation}
p(i, l, t) = [1+\mathrm{e}^{-\beta (\pi_l(t) -\pi_i(t))}]^{-1},
\label{imi_prob}
\end{equation}
where $\beta \geq 0$ controls the intensity of selection. Eq.~\eqref{imi_prob} is also used in Eq.~\eqref{phi} for the calculation of AGoS. For $\beta=0$, there is no selection pressure, meaning that evolutionary dynamics proceeds by random drift. As $\beta$ becomes larger, the tendency that strategies with higher payoffs will be imitated increases.

In the previous studies \cite{PinheiroPachecoSantos2012, PinheiroSantosPacheco2012}, the next time step immediately follows after the strategy updating.
In contrast, we incorporate mutation in this paper. Specifically, the strategy of individual $i$ changes to the different strategy with probability $m$ after the strategy updating.
We focus on how mutation alters the fitness of cooperation in networks.

\section{Results}
In each simulation, we calculated the GoS and conducted the strategy updating with mutation every time step, which was iterated $\Lambda=10^5$ time steps.
The population was composed of $N=1000$ (except for Lattice for which $N=31^2=961$) individuals and an average degree was $k=4$.
For each network type, we constructed 30 networks. For each network, we selected the number of initial cooperators from $j=0$ to $j=N$ randomly and this was repeated 1000 times.
Thus, we ran $\Omega=30 \times 1000=30000$ simulations in total for each network type and each mutation probability.
We varied mutation probability $m$ as the main experimental parameter.  
In the AGoS results, we obtained the locations of equilibrium points by numerically finding where the AGoS curve becomes zero. By doing so, the AGoS curves were smoothed using Gaussian kernels to reduce the effects of stochasticity in the results.

\subsection*{Case of accumulated payoffs}
\begin{figure*}[htbp]
\begin{center}
\includegraphics[width=3in]{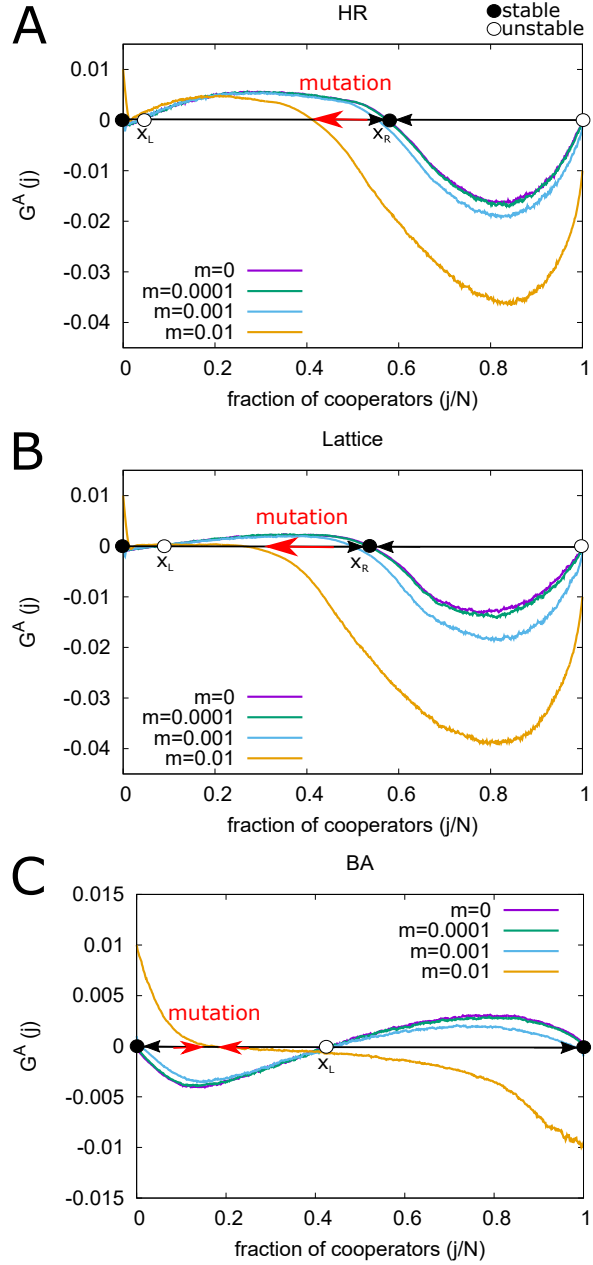}
\caption{AGoS with the accumulated payoffs. In all networks, mutation lowers the stable equilibrium points. (A) Homogeneous random networks (HR). $T=1.005$ and $\beta=10.0$. HR has two unstable equilibrium points ($x_L$ and 1) and two stable equilibrium points ($x_R$ and 0) at $m=0$.
(B) Square lattice (Lattice). $T=1.005$ and $\beta=10.0$. Similar to HR, Lattice has two unstable equilibrium points ($x_L$ and 1) and two stable equilibrium points ($x_R$ and 0) at $m=0$.
(C) BA scale-free networks (BA). $T=1.25$ and $\beta=0.1$. Unlike homogeneous networks (HR and Lattice), BA has one unstable equilibrium point ($x_L$) and two stable equilibrium points (0 and 1) at $m=0$.}
\label{AGoS_Acc}
\end{center}
\end{figure*}

We first see the results of accumulated payoffs.
Figure \ref{AGoS_Acc}A shows the AGoS ($G^A (j)$) on HR networks where the mutation probabilities $m$ are varied.
When $m=0$, the result perfectly matches the corresponding case of Ref.~\cite{PinheiroPachecoSantos2012}.
It has two stable equilibrium points and two unstable equilibrium points.
One of the unstable points, $x_L$, exists at $j/N =0.043$. One of the stable points, $x_R$, exists at $j/N=0.578$.
The other stable and unstable points exist at $j/N =0$ and $j/N =1$, respectively.
As the black arrows suggest, as long as $j/N > x_L$, the population composition converges to the stable point $x_R$.
Thus, unlike the case of well-mixed populations where defectors are dominant, cooperation and defection can co-exist in the network.
In other words, from a global, population-level perspective, HR networks can sustain cooperation even though all individuals play PD.

As $m$ becomes larger, $G^A (j)$ decreases overall. Thus, the stable coexistence point $x_R$ becomes lower (see the red arrow in Fig.~\ref{AGoS_Acc}A).
This means that mutation is harmful for the evolution of cooperation.
Mutation itself is neutral because both strategy changes (cooperation to defection and vice versa) take place without any bias.
Nevertheless, mutation negatively affects cooperation because it easily destroys the clusters of cooperators that allow cooperation to evolve.

At $j/N=0$, $G^A (j)$ naturally increases when mutation is considered.
At this point, $G^A (j)$ is the same as $m$, where there are only defectors and therefore any change from defector to cooperator must occur simply by mutation.

\begin{figure*}[htbp]
\begin{center}
\includegraphics[width=5in]{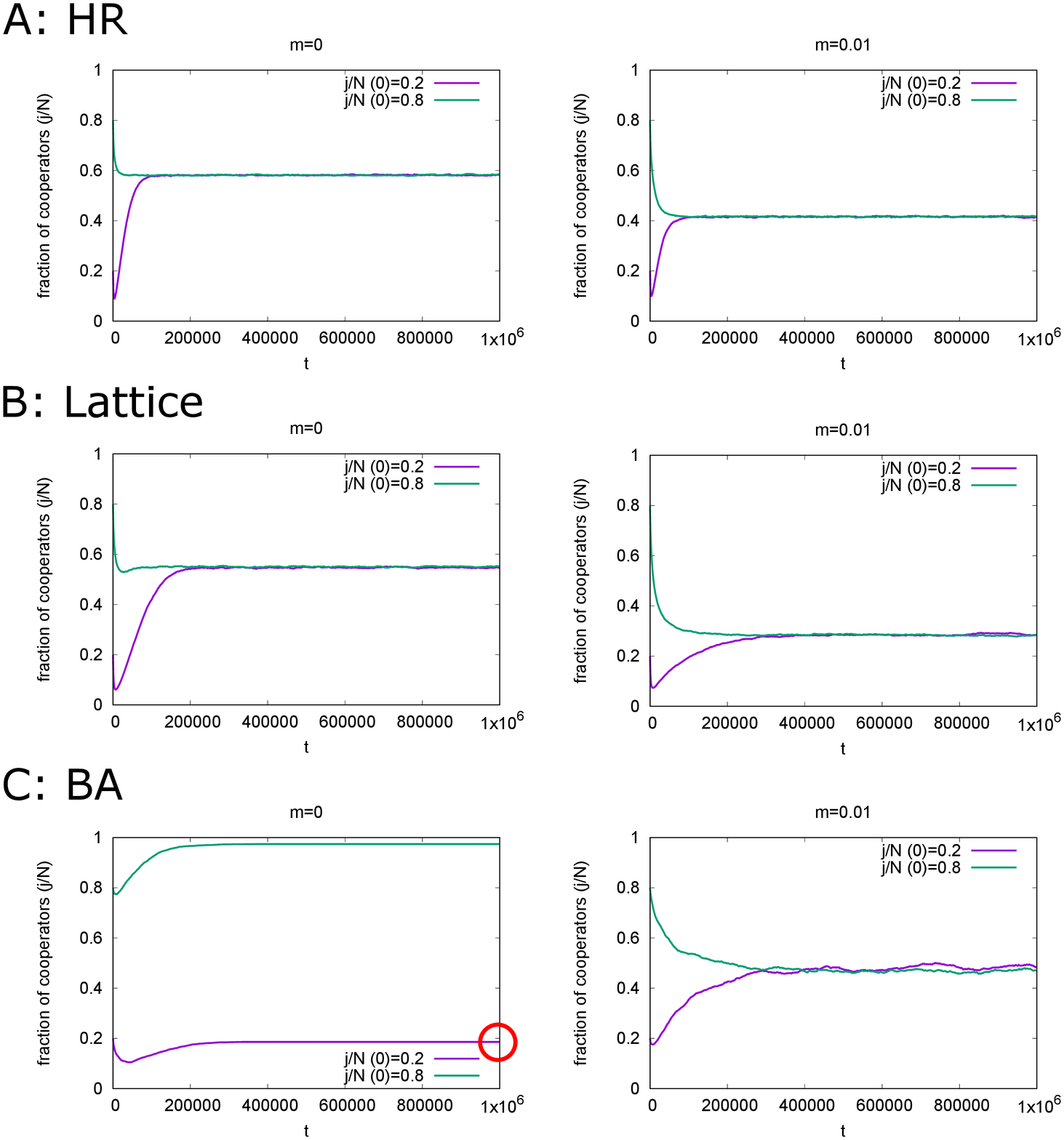}
\caption{Simulations of the strategy evolution on each network (A) HR, (B) Lattice, and (C) BA. The simulation results in HR and Lattice converged to each equilibrium point characterized by AGoS. Those results in BA have some deviations from the equilibrium points. $j/N (0)=0.2$ and $j/N (0)=0.8$ are used as the initial fractions of cooperators. Five hundred simulations are averaged. A million time steps are repeated for each simulation. Left: without mutation $m=0$. Right: with mutation $m=0.01$.}
\label{sim_Acc}
\end{center}
\end{figure*}

To verify the accuracy of our defined AGoS, we conducted simulations of the strategy evolution (Fig.~\ref{sim_Acc}A).
The results of five hundred simulations are averaged for each.
One million time steps are repeated for each simulation.
We used two different initial fractions of cooperators, $j/N (0)=0.2$ and $j/N (0)=0.8$, to check those converging points.
Without mutation ($m=0$; left), the fraction of cooperators converges to 0.58 irrespective of the starting points.
With mutation ($m=0.01$; right), it converges to 0.41 irrespective of the starting points.
We find that both converging points are nearly equal to the stable equilibrium points obtained through the analysis of AGoS.
Thus, AGoS is appropriate to characterize the fate of cooperation in HR networks.

Next, we discuss the case of lattice networks.
Figure \ref{AGoS_Acc}B shows  the AGoS on lattice networks where the mutation probability $m$ are varied.
The tendency of this result is the same with the HR networks because lattice is classified into homogeneous networks {(i.e., each node has the same degree).
When $m=0$, it has two stable equilibrium points and two unstable equilibrium points.
One of the unstable points, $x_L$, exists at $j/N =0.083$. One of the stable points, $x_R$, exists at $j/N=0.538$.
Hence, the density of cooperators in the coexistence situations is lower than HR networks ($0.538<0.578$).
The other stable and unstable points exist at $j/N =0$ and $j/N =1$, respectively.
If $j/N > x_L$, cooperation and defection can coexist in the network.
As $m$ becomes larger, $G^A (j)$ decreases overall. Thus, the stable coexistence point $x_R$ becomes lower.
Mutation negatively effects the evolution of cooperation in lattice networks, too.

Figure \ref{sim_Acc}B shows the simulations of the strategy evolution.
The setting is the same as HR networks. We used two different initial fractions of cooperators, $j/N (0) \approx 0.2$ and $j/N (0) \approx 0.8$ (these are approximate values because $N=961$ for Lattice), to check those converging points, respectively.
Without mutation ($m=0$; left), the fraction of cooperators converges to 0.55 irrespective of the starting points.
With mutation ($m=0.01$; right), it converges to 0.28 irrespective of the stating points.
We find that both converging points are nearly equal to the stable equilibrium points in the analysis of AGoS.
Thus, AGoS is appropriate to characterize the fate of cooperation in lattice networks.
Compared to HR networks every stable equilibrium point is lower in lattice networks.
In lattice networks, cooperative clusters tend to be localized and separated from each other.
It may prevent cooperators from expanding their regions.

We next see the results of BA networks.
Figure \ref{AGoS_Acc}C shows  the AGoS on BA networks where the mutation probabilities $m$ are varied.
When $m=0$, there is one unstable equilibrium point $x_L=0.402$ and two stable equilibrium points, $j/N=0$ and $j/N=1$.
This is the same as the one observed in the previous study \cite{PinheiroPachecoSantos2012}.
This means that cooperation becomes dominant once $j/N > x_L$.
BA networks with accumulated payoffs changed the game from defector dominance to a bistable one.
As $m$ increases, the number of cooperators needed for sustaining the dominance of cooperation becomes larger.
Interestingly, we found that, when $m=0.01$, the dynamics completely change from bistability to coexistence.
There exists only one stable equilibrium point at $j/N=0.142$. In other words, a relatively small population of cooperators will coexist with the majority of defectors in this case.

Figure \ref{sim_Acc}C shows sample simulation results of the strategy evolution.
The setting is the same with HR and Lattice networks.
Without mutation ($m=0$; left), the fraction of cooperators converges to 0.186 when $j/N (0)=0.2$ while 0.974 when $j/N (0)=0.8$.
Thus, there are some deviations from the corresponding analytical values, 0 and 1, which are characterized by AGoS.
In particular, the difference between 0.186 (see the red circle in the left panel of Fig.~\ref{sim_Acc}C) in the simulation and the value characterized by the AGoS (0).
Similarly, with mutation ($m=0.01$; right), both lines ($j/N (0)=0.2$ and $j/N (0)=0.8$) converge to around 0.475 in the simulations but the stable point characterized by AGoS is 0.142. Thus, there is also quite a difference between simulation and AGoS.

To clarify the reason, we examined individual simulation results.
Figure \ref{BaAccM0} shows all 500 runs without mutation ($m=0$) when $j/N (0)=0.2$.
From these results, we found that the fraction of cooperators did not converge to 0.186.
Instead, four hundreds and seven simulations converged to 0 while ninety three simulations converged to 1 in 500 total runs.
Thus, AGoS correctly characterized the final fate of cooperation with probability 0.814 but failed with probability 0.186.
In BA networks, degree-strategy correlations significantly affect the increase and decrease of cooperation.
However, the information cannot be captured by AGoS.
This is why AGoS sometimes fails to characterize the final fate of cooperation in BA networks.

\begin{figure}[htbp]
\begin{center}
\includegraphics[width=4in]{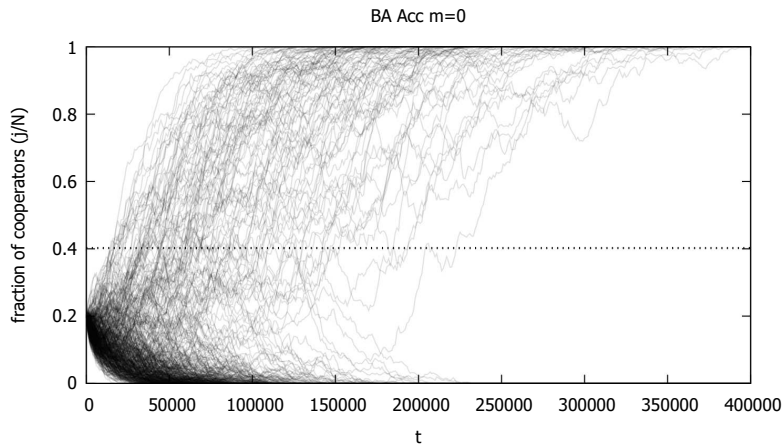}
\caption{All 500 simulation runs without mutation ($m=0$) when $j/N (0)=0.2$ on BA networks. Each line is almost transparent.
Four hundred and seven simulations converge to 0, which is the stable equilibrium point when it starts from $j/N(0)=0.2$ as characterized by AGoS. The color is dense because many lines are aggregated.
In contrast, ninety-three simulations converge to the opposite limit, 1. These rare cases are the cause of the deviation from the equilibrium point in BA. The dashed line denotes the internal unstable equilibrium point ($j/N =0.402$) without mutation.}
\label{BaAccM0}
\end{center}
\end{figure}

\subsection*{Case of average payoffs}
We now focus on how cooperation evolves with the average payoff compared to the accumulated payoff. It is known that, in networked games, cooperation is inhibited in the case of average payoffs because the hub effect is dismissed \cite{SantosPacheco2006, Szolnoki_etal2008, Tomassini_etal2007, Zhi-Xi_etal2007, IchinoseSayama2017}.
Figure \ref{AGoS_Ave} shows the case of average payoffs.
In HR networks (Fig.~\ref{AGoS_Ave}A), the internal stable equilibrium point is 0.598 without mutation ($m=0$) and is  0.437 with mutation ($m=0.01$). The corresponding simulations shown in Fig.~\ref{sim_Ave}A are 0.6 without mutation ($m=0$; left)  and are 0.44  ($m=0.01$; right), which means that the characterization by AGoS is quite accurate. 
In Lattice networks (Fig.~\ref{AGoS_Ave}B), the internal stable equilibrium point is 0.560 without mutation ($m=0$) and is  0.303 with mutation ($m=0.01$). The corresponding simulations shown in Fig.~\ref{sim_Ave}B are 0.55 without mutation ($m=0$; left) and are 0.28  ($m=0.01$; right).
The AGoS accurately characterizes the results of simulations in Lattice networks as well as the HR networks.

\begin{figure*}[htbp]
\begin{center}
\includegraphics[width=3in]{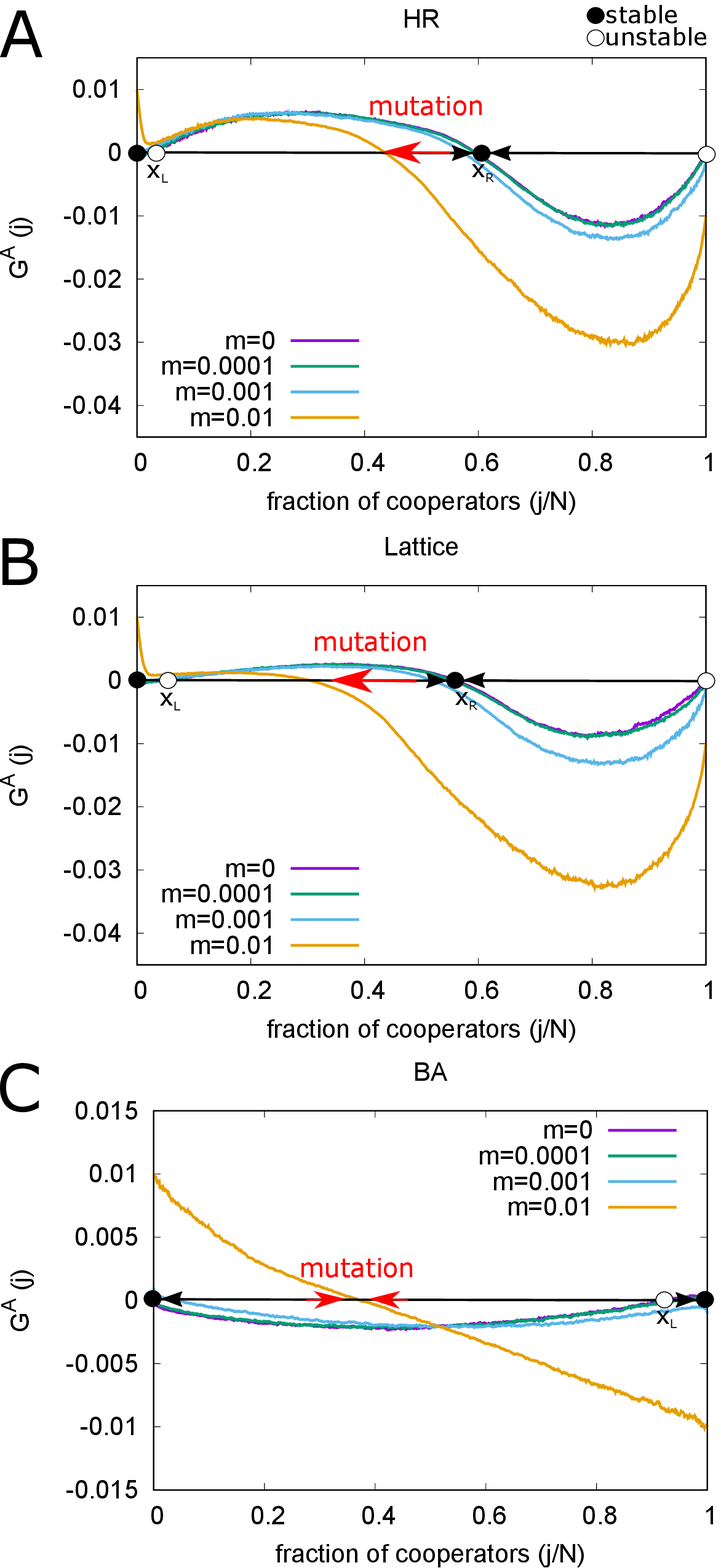}
\caption{AGoS on (A) HR ($T=1.005$ and $\beta=10.0$), (B) lattice ($T=1.005$ and $\beta=10.0$), and (C) BA networks ($T=1.25$ and $\beta=0.1$) with average payoffs. Mutation lowers the equilibrium points as well as in the case with the accumulated payoffs.}
\label{AGoS_Ave}
\end{center}
\end{figure*}

\begin{figure*}[htbp]
\begin{center}
\includegraphics[width=5in]{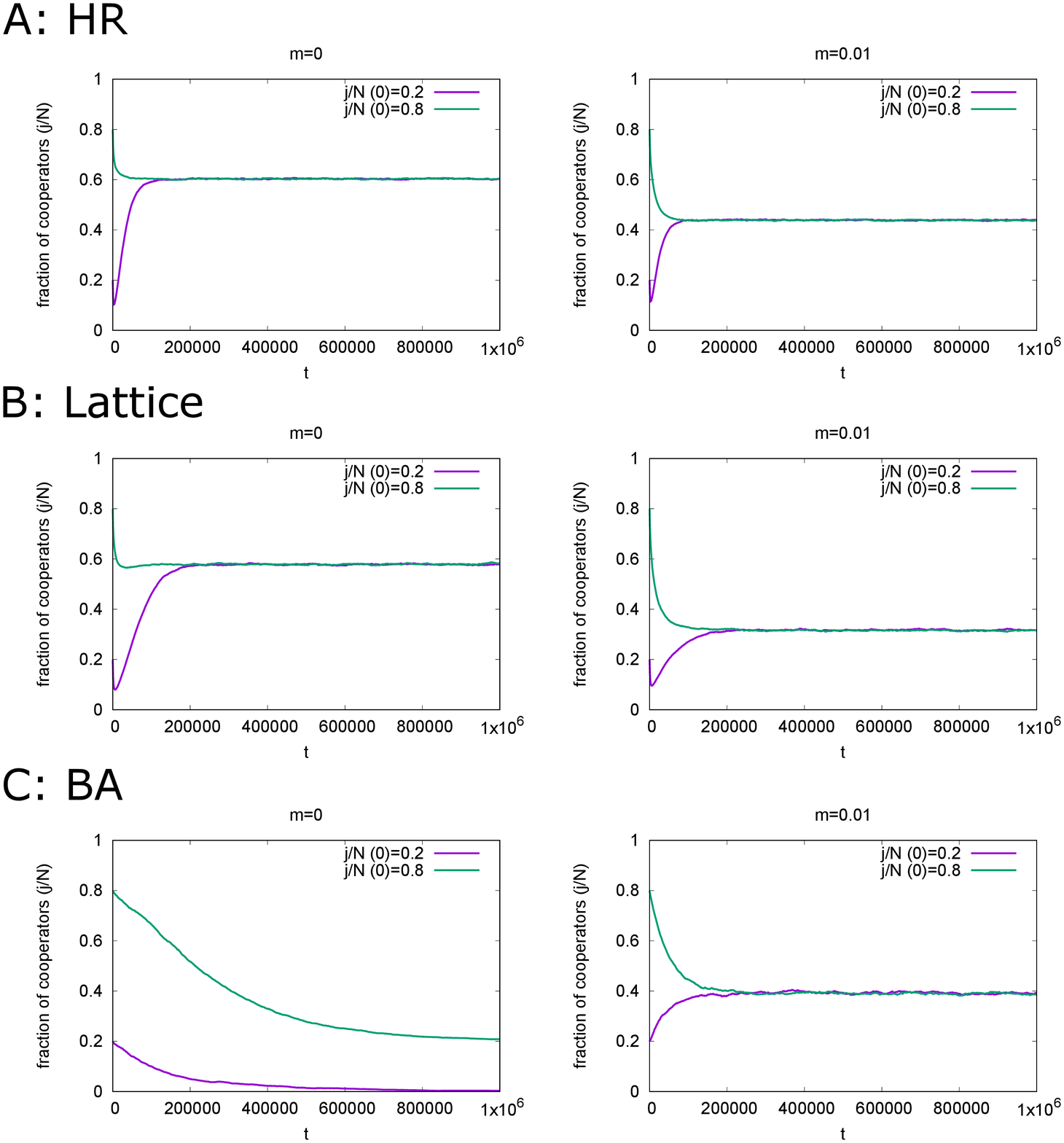}
\caption{Simulations of the strategy evolution on each network (A) HR, (B) Lattice, and (C) BA. The settings are the same as the accumulated payoffs. The simulation results (except for BA) converged to each equilibrium point characterized by AGoS.}
\label{sim_Ave}
\end{center}
\end{figure*}

For those two networks, if we compare the results of AGoS between accumulated (Fig.~\ref{AGoS_Acc}A and B) and average (Fig.~\ref{AGoS_Ave}A and B) payoffs, both equilibrium points in the average payoff are slightly higher than those in the accumulated payoff (see also Fig.~\ref{EquiPoints}).
There is little difference between the average and accumulated payoffs because each node has the same degree in those networks.
However, if the payoffs are averaged, the payoff difference between cooperators and defectors shrinks, hence the chance that cooperators can survive slightly increases in the average payoffs.

We move to the case of BA networks.
Figure \ref{AGoS_Ave}C shows  the AGoS on BA networks where the mutation probabilities $m$ are varied.
When $m=0$, there is one unstable equilibrium point $x_L=0.917$ and two stable equilibrium points, $j/N=0$ and $j/N=1$.
This means that cooperation becomes dominant if $j/N > x_L$.
Compared to the case of accumulated payoff (Fig.~\ref{AGoS_Acc}C), the critical point $x_L$ greatly moves rightward, which means that the critical fraction of cooperators needed for cooperation to become dominant is quite large.
This is because the hub effect is dismissed in the case of average payoffs.

The corresponding simulations (Fig.~\ref{sim_Ave}C) show that AGoS characterizes accurately when $j/N(0)=0.2$ where the simulations converge to 0 but largely fails when $j/N(0)=0.8$ where the the population converges to full cooperation 20\% of the time and full defection 80\% of the time, resulting in around 0.2.
The reason is that degree-strategy correlations are missed in AGoS in BA networks.

When $m=0.01$, the dynamics changes from bistability to coexistence in the same way as with the accumulated payoffs.
The stable equilibrium point is 0.372 (Fig.~\ref{AGoS_Ave}C) and the fraction of cooperators converges to 0.39 in the corresponding simulations (right in Fig.~\ref{sim_Ave}C).
We consider the possible reason for the change in these dynamics in the next subsection.

\subsection*{Mutation effect on BA networks}
We saw that the dynamics completely changed on BA networks in both accumulated and average payoffs.
To consider which factor contributes to the change of dynamics, we show a case in which only mutation occurs with probability 0.01 ($m=0.01$) without strategy updating (red line in Fig.~\ref{MutationBA}).
If there is no cooperator ($j/N=0$), the expected probability that cooperator increases by one is 0.01 ($G^A(j)=0.01$). In contrast, if there is no defector ($j/N=1$), the probability that cooperator decreases by one is 0.01 ($G^A(j)=-0.01$). If cooperators and defectors are fifty-fifty, that probability becomes 0.
Note that, in this case, because there is no strategy updating these results always hold irrespective of network structures because the locations of individuals and their payoffs do not matter.

We first see the comparison of the accumulated payoffs with the case of only mutation.
In Fig.~\ref{MutationBA}, it seems that the dynamics are heavily influenced by mutation because the solid orange line curves in accordance with the red line.
In other words, the dynamics changed from bistability to coexistence due to the effect of mutation.
In BA networks, contrary to HR or Lattice networks, once a cooperative hub is destroyed by mutation, there is almost no chance for cooperation to evolve.
This is why BA networks are most vulnerable to mutation compared to HR or Lattice networks.

As in the case of accumulated payoffs, the change of dynamics in the average payoffs arises from the effect of mutation.
As seen in Fig.~\ref{MutationBA}, the AGoS in the average payoff (dashed line) is closer to the case where only mutation is considered because there is no hub effect in this case.
Thus, the dynamics in the average payoffs are highly influenced by mutation compared to the case of the accumulated payoffs.

\begin{figure}[htbp]
\begin{center}
\includegraphics[width=4in]{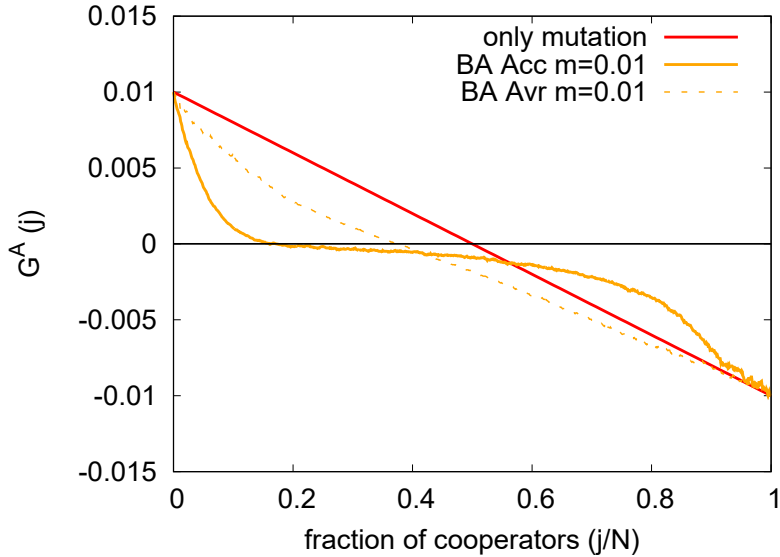}
\caption{Comparison between the accumulated (orange solid) and average (orange dashed) payoffs on BA networks and the case of only mutation (red). This figure shows that mutation strongly affects the dynamics in BA networks. In the last case, only mutation occurs with probability 0.01 ($m=0.01$) without strategy updating. Note that, in this case, network structures and payoffs do not matter because one randomly selected individual changes its strategy with probability $m=0.01$.}
\label{MutationBA}
\end{center}
\end{figure}


\subsection*{Comparison of equilibrium points}
Finally, we take a close look at the location of stable equilibrium points against mutation probability on the three networks.
Here, $m$ is varied from 0 to 0.1.
Figure \ref{EquiPoints} shows the results in (A) HR, (B) Lattice, and (C) BA networks.
In each network, the solid lines indicate the stable equilibrium points and the dashed lines indicate the unstable equilibrium points.
Blue (Red) is the case of the accumulated (average) payoffs.

By comparing (A) with (B), we find that the stable equilibrium points in HR networks are always higher than those in Lattice networks.
This is because cooperators are often locally clustered in Lattice networks compared to HR networks as previously described.
In those two networks, as $m$ becomes larger, the stable equilibrium points decrease unless $m$ is too large (e.g. $m=0.1$).
The reason is that cooperative clusters are destroyed by mutation.
In contrast, if $m$ is too large, cooperators are continuously produced by mutation irrespective of the strategy updating.
It contributes to the rise of equilibrium points.


\begin{figure*}[htbp]
\begin{center}
\includegraphics[width=3in]{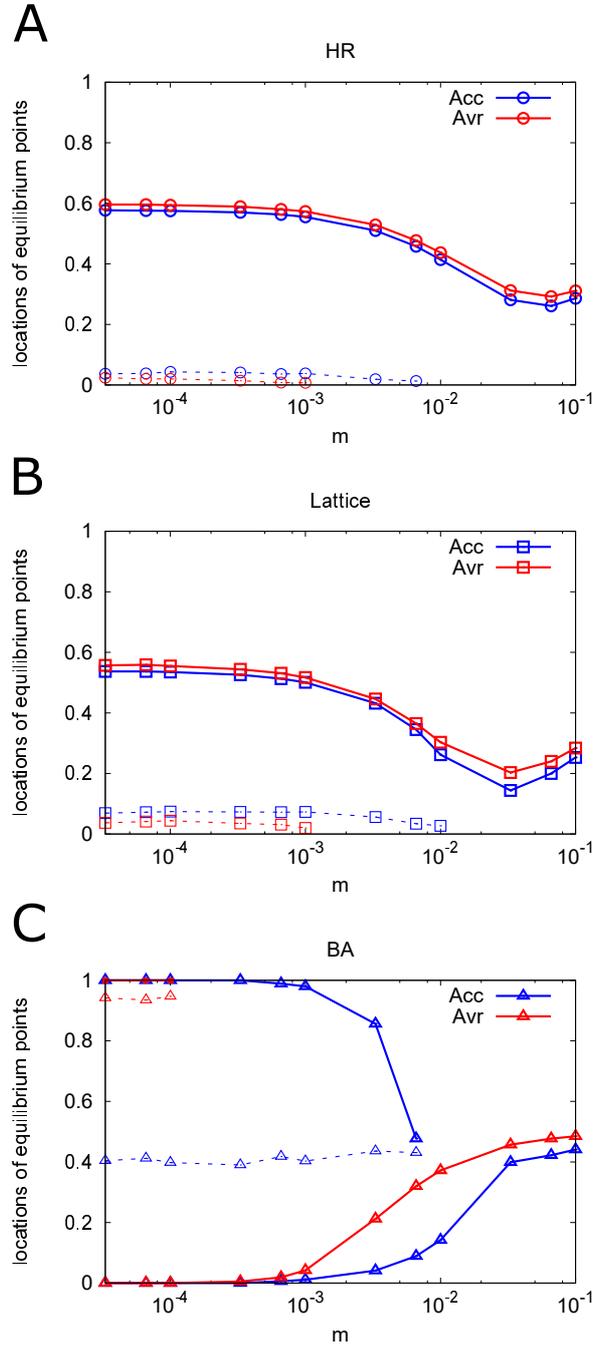}
\caption{Location of equilibrium points as a function of mutation probability $m$ on each network: (A) HR, (B) Lattice, and (C) BA.
In each network, the solid lines indicate the location of stable equilibrium points and the dashed lines indicate the location of unstable equilibrium points. Blue (Red) is the case of the accumulated (average) payoffs. The stable equilibrium points in HR networks are always higher than those in Lattice networks. Moreover, in BA networks, the dynamics change from bistability to coexistence.}
\label{EquiPoints}
\end{center}
\end{figure*}

In BA networks, as we previously mentioned, the dynamics change from bistability to coexistence.
Moreover, by changing $m$ in more detail, we find the imperfect pitchfork bifurcations in the accumulated payoffs (Fig.~\ref{EquiPoints}).
The main reason of the change to coexistence was that scale-free networks are vulnerable to mutation.
Once cooperative hubs (key for cooperation to evolve in scale-free networks) are destroyed by mutation, the dynamics are mostly governed by mutation, which lead to the coexistence.
Because there is no hub effect in the average payoffs, the coexistence takes place at smaller $m$ values.

\section{Discussion and conclusions}
In this paper, we analyzed the evolution of cooperation at a population level in the presence of mutation by the AGoS.
We used two classes of networks: homogeneous (HR and Lattice) and heterogeneous (BA).
We also examined two different payoff mechanisms: accumulated and average payoffs.
Our analyses revealed that, as the prominent mathematical studies showed \cite{Allen_etal2012, Debarre_etal2017}, mutation has a negative effect on the evolution of cooperation regardless of the payoff functions, fraction of cooperators, and network structures, because clusters of cooperators can easily be destroyed by mutation.
Moreover, we found that scale-free networks are most vulnerable to mutation because a few cooperative hubs are the only key of the success for cooperation, despite the fact that cooperation is greatly enhanced with accumulated payoffs without mutation in those networks. If a few hubs are destroyed by mutation, cooperation easily collapses.
This may be relevant to the robustness of  scale-free networks. In general, scale-free networks are quite weak when hubs are destroyed by targeted attacks \cite{Albert_etal2000}.
Moreover, cooperation on scale-free networks easily collapses when hubs are removed \cite{Perc2009, Ichinose_etal2013}.
Our paper shows that, even if hubs are not directly removed, scale-free networks are vulnerable to mutation because hubs are easily invaded by defectors through mutation.
Therefore, to prevent cooperative societies from collapsing, some mechanisms may need to be developed regarding how hubs are protected from mutation.

Similar to recent mathematical results \cite{Allen_etal2012, Allen_etal2017, AllenNowak2014, Tarnita_etal2014, Debarre_etal2017}, our results also indicate the importance of considering random noise (mutation), which was largely overlooked in the literature, in studying the evolution of cooperative behavior in social networks.
Mutation can be considered genetic change if we assume biological systems. If we assume cultural systems, mutation can be considered a stochastic behavior to explore new behaviors.
Although we showed such a random exploration was harmful to maintain cooperative societies, if we consider different forms of exploration, those explorations may work beneficially for societies, such as collaborative problem solving \cite{SayamaDionne2015}.
We would like to consider those cases in the future work.

Finally, we remark on the limitations of this paper. Our extended AGoS considering mutation sometimes fails to characterize the final states of cooperation on BA networks, although it works perfectly for HR and Lattice networks.
This was because AGoS aggregate every state and topology of networks into one value. To consider those complicated network interactions, we may need to take into account the influence of nodes separated by the degree.
A similar idea is the reproductive value, which is a measure of the expected gene contribution of an individual to the future gene pool \cite{Fisher1930, Maciejewski2014, TarnitaTaylor2014, Allen_etal2017}. Even if there is no selective difference among all individuals, individuals whose degree is high may be favored by natural selection more often than the others on networks. This is the reproductive value based on degree difference \cite{Maciejewski2014}.
In this study, we could elaborate our extended AGoS by considering such reproductive values based on degree difference, but it is not our main focus. Rather, we would like to emphasize that the change of dynamics in heterogeneous networks can be clearly shown even by our ``rough" AGoS. The elaboration of the AGoS would be an important future direction.

\section*{Competing interests}
We have no competing interests.

\section*{Author's contributions}
GI and HS designed the research. GI and YS developed the model. GI and YS performed the simulation.
GI and HS discussed and analyzed the results. GI and HS wrote the main manuscript text. All authors gave final approval for publication.

\section*{Acknowledgment}
G.I. acknowledges the support by Hayao Nakayama Foundation For Science \& Technology \& Culture.


\end{document}